# Real-time video streaming *in vivo* using ultrasound as the communication channel

Zhengchang Kou, Rita J. Miller, Andrew C. Singer, *Fellow, IEEE*, and Michael L. Oelze*, *Senior Member, IEEE*

*Abstract*—The emergence of capsule endoscopy has provided a means of capturing video of the small intestines without having to resort to an invasive procedure involving intubation. However, real-time video streaming to a receiver outside the body remains challenging for capsule endoscopy. Traditional electromagnetic-based solutions are limited in their data rates and available power. Recently, ultrasound was investigated as a communication channel for through-tissue data transmission. To achieve real-time video streaming through tissue, data rates of ultrasound need to exceed 1 Mbps. In a previous study, we demonstrated ultrasound communications with data rates greater than 30 Mbps with two focused ultrasound transducers using a large footprint laboratory system through slabs of lossy tissues [1]. While the form factor of the transmitter is also crucial for capsule endoscopy, it is obvious that a large, focused transducer cannot fit within the size of a capsule. Several other challenges for achieving high-speed ultrasonic communication through tissue include strong reflections leading to multipath effects and attenuation. In this work, we demonstrate ultrasonic video communications using a mm-scale microcrystal transmitter with video streaming supplied by a camera connected to a Field Programmable Gate Array (FPGA). The signals were transmitted through a tissue-mimicking phantom and through the abdomen of a rabbit *in vivo*. The ultrasound signal was recorded by an array probe connected to a Verasonics Vantage system and decoded back to video. To improve the received signal quality, we combined the signal from multiple channels of the array probe. Orthogonal frequency division multiplexing (OFDM) modulation was used to reduce the receiver complexity under a strong multipath environment.

## I. INTRODUCTION

Capsule endoscopy is a technology that was introduced two decades ago by Given Imaging [2][3][4]. Within two years of its introduction, capsule endoscopy was approved for use in the USA and Europe [5][6]. Capsule endoscopy is employed clinically to provide diagnostic information for diseases of the stomach and small intestines. This approach is less invasive and is, therefore, preferred, when possible, over traditional methods where a scope on a cable is traversed down the esophagus and into the stomach and small intestines. Furthermore, not all of the small intestines are accessible via scopes due to its many loops and turns. Capsule endoscopy systems, i.e., camera pills, have been created to wirelessly transmit image data through the human body as the pill traverses through the small intestines. Currently, all wireless camera pills transmit data via an electromagnetic (EM) communication modality to a receiving antenna outside the body, where the signal is decoded, and images stored for later viewing.

Traditional EM-based wireless communication approaches face several limitations for in-body communication, especially in the hospital environment. First, the high attenuation of EM waves inside the human body can result in low signal-to-noise ratio (SNR) and limited penetration depth. Second, the power supply of the capsule is limited because the batteries must be small to fit in the capsule, which limits the transmit power. Third, the temperature rise caused by the absorption of EM energy by tissues also limits the transmit power. Because of these limitations, capsule endoscopy using EM wave communication in the human body has not achieved data rates capable of real-time video streaming.

To address some of these issues, in one study [7] a conformal helical antenna was investigated as a transmitter antenna to reduce the elevation in temperature resulting from EM radiation. The authors demonstrated a data rate of 256 kbps with a low error rate ($<10^{-3}$) in an *in vivo* test with specific absorption rate well below FCC recommendation. In other work [8], a backscatter communication approach was used to move all the communication power consumption outside of the body. In that study, a data rate of up to 1 Mbps was achieved over an 8-cm distance at 0 dBm transmitting power. However, the system occupied the 25 MHz RF bandwidth, which has the potential to interfere with other equipment using similar bands and violates current FCC regulations.

As an alternative to EM communications, ultrasound can provide a communication channel in the human body that has the potential to provide sufficiently high data rates to stream real-time high-definition (HD) video. Because it is ultrasound, it will not interfere with other equipment in the hospital setting. In a previous study, we demonstrated high-speed communication using ultrasound through tissue [1], with data rates of up to 30 Mbps achieved through 5-cm thick pork loin and beef liver samples. The transducer used in the study had a wide bandwidth (-10-dB bandwidth of 5 MHz), a center frequency of 4 MHz and large diameter (1.92 cm). In that system, the communication waveforms were generated by an arbitrary waveform generator connected to a power amplifier that was then connected to the transmitting transducer. The matched receiving transducer was connected to a computer for processing. This laboratory system was used to demonstrate the capability of ultrasound to transmit high data rates through tissues sufficient for streaming video. However, the footprint of the laboratory system required a small table to house it and, therefore, would not be small enough to fit inside a camera pill.

To enable in-body ultrasound communications with a footprint small enough to fit in an endoscopy capsule and with data rates capable of streaming video requires small transducer elements, a small processing unit and sufficient ultrasonic power. In a recent study by Bos, et al., small individual ultrasonic elements, i.e., the Sonometrics microcrystal (Sonometrics, London Ontario), were used as transmit and receive transducers [9], [10]. These microcrystals are 2 mm in diameter, operate at center frequency of 1.2 MHz, are biocompatible and have low directivity. In one of those studies, data rates of 4.4 Mbps were achieved through beef liver with quadrature amplitude modulation (QAM) and a decision feedback equalizer [10]. Bos et al. used QAM and orthogonal frequency division multiplexing (OFDM) modulation to achieve data rates over 300 kbps through beef samples and over distance of 10 cm [11]. These studies demonstrated that using ultrasound as a communication medium with OFDM was feasible for in-body communication with small footprint transducers. However, the data rates achieved were insufficient for real-time video transmission.

In studies by Santagati et al. [12], [13], an IoT network using ultrasound was implemented. In that work, they used a customized 9.5-mm diameter thin disk piezoelectric transducer and pulse-position modulation (PPM). The processing system consisted of a small footprint FPGA and microcontroller. A physical layer was implemented on an FPGA while a media access control layer was positioned on a microcontroller unit to maximize battery life. The use of PPM also simplified the transmission because there was no need to use an external DAC. The network they built was suitable for low speed data or control signal transmission. They achieved data rates of 180 kbps at -20-dBm transmit power. However, they did not provide video capable data rates using ultrasound with a small footprint processing unit.

In this study, we demonstrate streaming of real-time video using ultrasound as a communication modality, an ultrasonic array connected to a commercial ultrasonic scanner to capture the ultrasonic signals and a small form factor source transmitting signals using a small footprint FPGA board. The signals were received and processed using a commercial ultrasound scanner and array probe because there is less of a constraint for the size of the external receiving device. This work moves toward miniaturization of the transmission source for use in capsule endoscopy and how these signals might be captured for processing by an external array receiver. In this paper, Section II first describes the channel model we used to simulate the in-body communication environment and then describes both the methodology employed for the modulation and equalization schemes, the test setup and FPGA implementation of transmitter. Section III provides the results of the study for both the simulation and phantom-based test and images received in an *in vivo* test. The last section (Section IV) provides discussion and conclusions regarding the study.

## II. Methodology

### A. Ultrasound Channel Model

To model and capture the characteristics of an in-body ultrasonic communication channel, i.e., to estimate what the delay spread due to multipath might be in a tissue region like the human abdomen, a small tissue-mimicking phantom with bones or other reflectors inside can be constructed and the finite-length impulse response (FIR) channel can be measured between transducers embedded inside the phantom and external receivers [9]. Here we constructed a cylindrical-shaped phantom to partially mimic the human abdomen. The phantom was 10 cm in diameter and 9 cm in height and was made with a 1.5% agar and 1.5% gelatin mixture as shown in Fig. 1. A microcrystal transducer (Sonometrics, London Ontario) was used as the transmitter. The microcrystal was unfocused, has a 2-mm diameter and a nominal center frequency of 1.2 MHz. The microcrystal was embedded in the phantom 5 cm from the top and 6.5 cm from the edge of the phantom. In this experiment, an IP103 64-element phased array (Sonic Concepts, Bothell, WA) was used as the receiving array positioned along the side of the phantom. The array had a center frequency of 3 MHz with an element height of 14 mm and pitch of 0.304 mm.

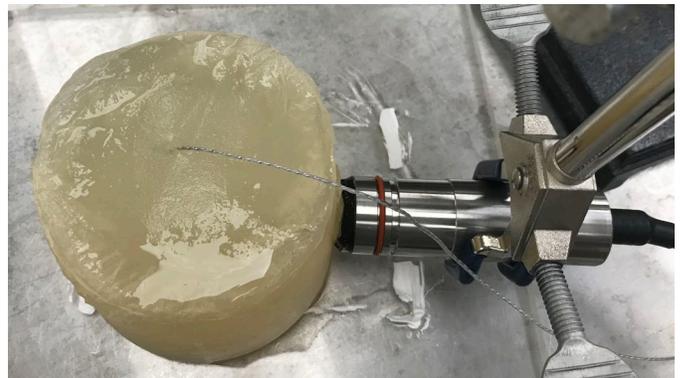

Fig. 1. IP103 probe positioned along the side of the phantom. The cable attached to the microcrystal can be observed penetrating into the phantom at its center.

We used estimated the FIR channel by generating a symbol sequence in Matlab (MathWorks, Natick, MA), loading the sequence into a Tabor WW1281 arbitrary waveform generator (AWG) (Tabor Electronics, Haifa, Israel), transmitting the sequence with the microcrystal and recording the received signal with the IP103 64-element phased array (Sonic Concepts, Bothell, WA) connected to a Verasonics Vantage 128 system (Verasonics, Kirkland, WA). The FIR channel measured by the center element of the probe is shown in Fig. 2. As the size of phantom was relatively large compared to a wavelength and strong reflections occurred at the boundary between the air and phantom material, delay spread up to 1 ms was observed. To minimize the performance loss due to the multipath delay spread, the symbol duration was chosen to be much longer than the delay spread, which results in a narrow subcarrier bandwidth. The frequency response of the channel was estimated over the occupied ultrasonic frequency band measured by all 64 elements.

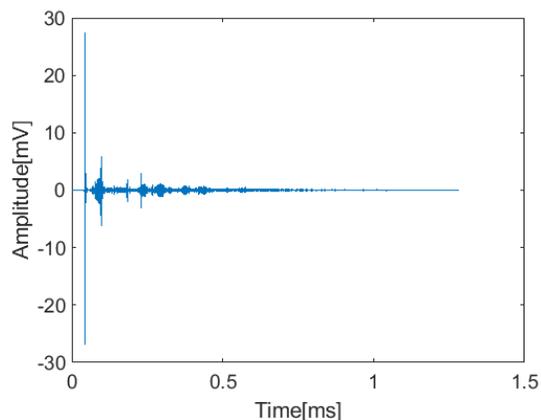

Fig. 2. FIR channel measured with the center element of receiving array. Delay spread can be up to 1 ms long.

### B. Modulation and Equalization scheme

According to the channel model measured from the phantom, it was observed that the delay spread was up to 1 ms, which results in a narrow coherence bandwidth. To mitigate this, we used OFDM modulation with a narrow subcarrier bandwidth to simplify the equalization process. QAM signaling is used in each subcarrier. The proposed OFDM frame structure is composed of pilot symbols and data symbols. Figure 3 shows the details of the frame structure, also called a block type pilot [14].

The number of data symbols between two pilot symbols was determined by the rate of channel variation. We used two pilot symbols to estimate the channel and interpolated through these data symbols to perform equalization. A pseudo noise (PN) sequence was used as the pilot signal. A block-type pilot sequence provided a higher channel estimation accuracy when the coherence bandwidth was narrow, and the channel variation rate was low compared to the distance between two pilot symbols. This is because we had the channel measurement at each individual subcarrier, compared to previous studies, which had pilot tones distributed amongst the subcarriers [11], [15].

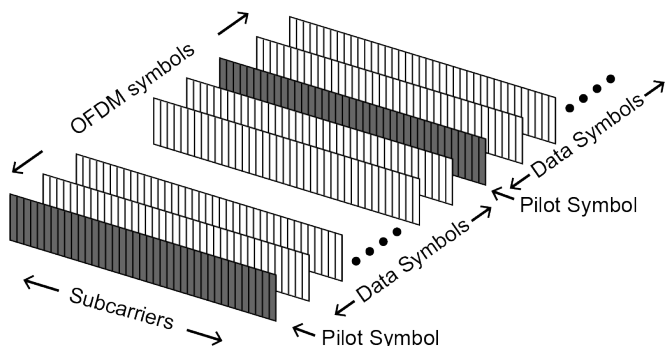

Fig. 3. Frame structure of OFDM modulation used in this study. Shaded are pilot symbols, unshaded symbols between pilot symbols are data symbols.

To improve the signal-to-noise ratio (SNR), the receiving area was increased by using an ultrasonic array as the receive transducer. Passive receive focusing then was used to sum the signals from each channel to further boost the SNR. Compared to single-element focused transducers we could automatically focus on the transmit element with the array, reducing the requirement that the transmitter and receiver be perfectly aligned.

As Fig. 4 illustrates, there are path differences between the transmit microcrystal to different receive elements on the array, which results in different time delays to each of the receive channels. To passively focus on the transmit microcrystal we performed symbol timing recovery (STR) on each receive channel, which aligned all receive channels to the same time point up to the precision of sampling rate. The phase offsets caused by residual delay differences between channels were compensated by performing channel equalization on each channel's subcarrier.

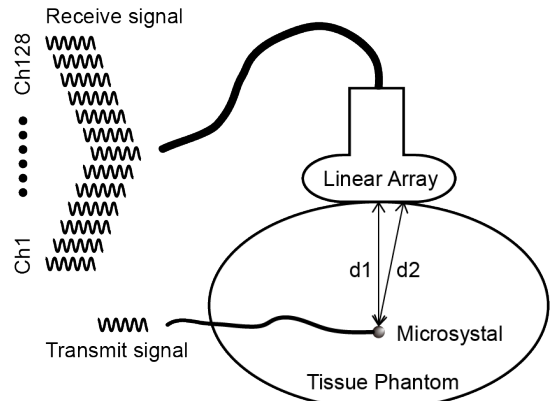

Fig. 4. Illustration of time delay difference between receive signals from different elements of the linear array.

The overall processing chain is displayed in Fig. 5. On each of the multiple receive channels the following operations were performed: digital down conversion (DDC), STR, Frequency shift compensation (FSC), Fast Fourier transform (FFT), Channel Estimation and Equalization (ChEst, ChEq) individually. At the final stage, the product of all channels was combined together by maximum ratio combining (MRC) [16].

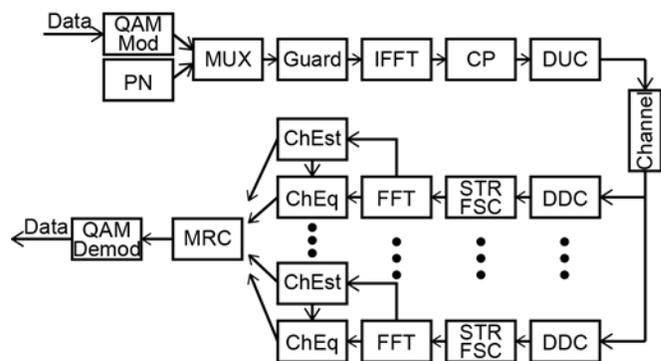

Fig. 5. Structure of OFDM modulation and demodulation used in this study.

To mitigate inter symbol interference, a cyclic prefix (CP) based symbol timing recovery was used with the frequency shift estimation method [17]. After applying frequency offset estimation, we can correct for frequency offset and the CP symbol timing recovery provides the starting point of each OFDM symbol, from which we can set the start point of the FFT to perform the OFDM demodulation. Because the pilot symbol is obtained at the beginning and ending of one frame, which provides the pilot signal at each subcarrier, we can estimate the channel by dividing the received pilot signal by the transmitted pilot signal [18]. The channel estimate is interpolated through the data symbols between the two pilot symbols.

The anticipated multipath environment in a medium like the human abdomen causes frequency selective fading, which results in poor SNR for some subcarriers. To alleviate this problem, we applied a weighted combination of a subcarrier from different receive channels with weights selected according to the channel response magnitude for the specific subcarrier, i.e., MRC, if the frequency selective fading does not impact all channels.

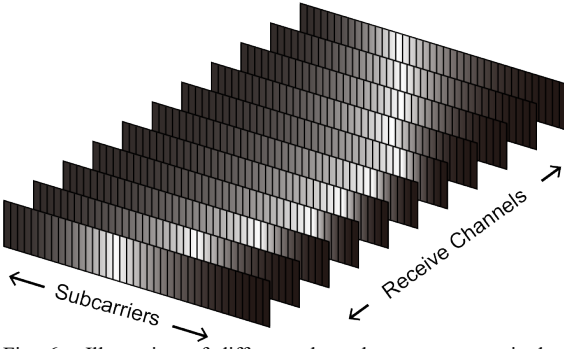

Fig. 6. Illustration of different channel response magnitudes in different channels and subcarriers, where the brightness reflects how deep the frequency selective fading is.

As Fig. 6 illustrates, the brighter subcarrier corresponds to a stronger frequency selective fading. Because the receive elements are spatially distributed, we can have different fading profiles in different channels. Here we define the number of channels as $N$, the number of subcarriers as $B$, the equalized signal at subcarrier $i$ and channel $j$ as $S_{i,j}, i \in [1, B] \ j \in [1, N]$ and the magnitude of the channel response at the same position is $H_{i,j}, i \in [1, B] \ j \in [1, N]$. Then, the MRC can be calculated as,

$$C_i = \frac{\sum_{j=1}^{N} S_{i,j} * H_{i,j}}{\sum_{j=1}^{N} H_{i,j}}. \tag{1}$$

After calculating and applying the MRC product, the processed signal is quantized to the QAM symbol map and then demodulated to recover the original transmitted data.

The modulation parameters used in this study are listed in Table I. A 4096-point FFT was chosen, as compared to [11], which used 64 and 32,768, respectively. A 64-point FFT was too small for our purpose as the subcarrier bandwidth would be too wide, while 32,768 was too large to implement on a resource limited FPGA. The total bandwidth used was 937.5 kHz. With 16 QAM modulation, the overhead of CP and pilots, the payload data rate was 3 Mbps, which is sufficient for HD video streaming using H.264 encoding. With 256 QAM modulation, a higher data rate of 6 Mbps was achievable. We also applied the same modulation scheme with a higher baseband sampling rate and center frequency to use a larger bandwidth, which enabled even higher data rates of 15.2 Mbps. The higher bandwidth of 2.3 MHz was more than twice the original bandwidth. In this study, we did not include any channel coding or video coding because the aim of the study was to demonstrate the capability of using ultrasound as a channel for video capable capsule endoscopy.

### C. Test Setup

Waveforms corresponding to each frame of data, along with the pilot symbols, were generated via Matlab using custom scripts and the waveforms were loaded into the AWG, which was connected to the microcrystal acting as the transmitter. To closer approach the voltage levels that might be achieved in a capsule endoscopy, we limited the highest transmit voltage to 1 V peak. For the receiver in the phantom experiments we used the IP103 phased linear array and a Verasonics Vantage 128 ultrasound imaging research platform. As the Verasonics system works in the Matlab environment, we could easily retrieve the radio frequency data and perform demodulation in a continuous way. The integrated analog front end of Verasonics provides over 40 dB in gain along with a programmable low pass filter. In this way, there was no need to use an external amplifier to amplify the weak signal from the array elements, which reduced the system complexity. Fig.7 provides the bandwidth of the impulse response from the signal transmitted by the microcrystal and received by the IP103 array.

TABLE I
MODULATION PARAMETERS

| Parameter | Symbol | Value | Parameter | Symbol | Value |
|---|---|---|---|---|---|
| FFT size | $N_{FFT}$ | 4096 | Sample Rate | $F_s$ | 1.25/3.13 MSPS |
| Active Carriers | $N_{AC}$ | 3072 | Modulation Depth | $M$ | 16/256 |
| Subcarrier Bandwidth | $\Delta_f$ | 305/763 Hz | Center Frequency | $F_C$ | 2.4/3.2 MHz |
| CP length | $T_{CP}$ | 409/164 µs | Block distance | $D_{pilot}$ | 12 |

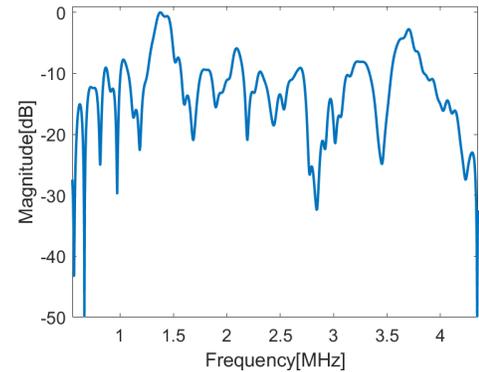

Fig. 7. Frequency response of impulse from the microcrystal received by IP103 array.

### D. FPGA Implementation

After successfully demonstrating the video streaming with our laboratory system and to move to a smaller hardware footprint, an FPGA board was used for producing the transmission signals. We used a Redpitaya STEM 125-14 as the hardware development platform for the transmitter. The platform uses a Xilinx Zynq 7010 FPGA and dual channel 125 MSPS 14-bit digital to analog converter (DAC) along with an onboard amplification circuit. The FPGA design is comprised of three major parts: camera interface, OFDM modulator and digital up converter. The camera interface was written in Verilog using Xilinx Vivado IDE. The OFDM modulator and digital up converter were designed and generated inside the Xilinx System Generator for digital signal processing (DSP). The modulation scheme described in part B was implemented on the FPGA. The digital up converter design was modified to fit into the limited DSP resources on the FPGA. This design could transmit a 160*50 RGB565 color image from an OV7670 VGA camera module for up to 30 frames per second. The resource utilization and power estimation of FPGA for this task are provided in Table II. The DAC provided two single ended outputs, which were connected to a single microcrystal differentially providing a maximum of 1 Vpp for transmission. We used the FPGA system with the ultrasonic microcrystal for video transmission through the phantom and for the *in vivo* experiments.

TABLE II
FPGA RESOURCE UTILIZATION AND POWER CONSUMPTION

| Parameter | Usage | Parameter | Usage |
|---|---|---|---|
| Look Up Table | 36% | DSP slice | 73% |
| Flip-Flop | 28% | Dynamic Power | 0.264 W |
| Block RAM | 73% | Static Power | 0.098 W |

### E. In vivo Setup

The protocol was approved by the Institutional Animal Care and Use Committee (IACUC) at the University of Illinois at Urbana-Champaign. A New Zealand White rabbit was used in the *in vivo* experiments. The rabbit was anesthetized with isoflurane during the procedure and euthanized after the procedure. During anesthesia, the rabbit was placed on its back and its abdomen was shaved and depilated. A small incision was made in the abdomen of the rabbit on its left side below the ribcage. The microcrystal, which was connected via cable to the FPGA, was inserted inside the abdomen approximately 2 cm from the outer skin surface. The FPGA was also connected to a small camera outside of the rabbit. An array probe (C5-2) connected to the Verasonics system was used to receive the signals through the rabbit abdomen. Ultrasonic gel was used to couple the array probe to the outside of the abdomen. Signals were transmitted via ultrasound from within the rabbit abdomen using the embedded microcrystal out to the array sitting on the outer surface a few cm away.

## III. RESULTS

### A. Phantom Results

By convolving the FIR measured with IP103's 64 receiving elements from the phantom with the passband signal, we were able to apply the channel to the up converted signal to perform the simulations over a realistic channel model provided by the phantom. The results are shown in Fig. 8. As Eb/N0, which is defined as the energy per bit over noise power, increased, the MRC results provided better bit-error-rate (BER) performance, which improves as the number of channels increases.

We transmitted the up converted OFDM signal through the phantom and recorded the received signals. By varying the maximum output voltage of the AWG from 50 mV to 1 V peak we achieved both 16 QAM and 256 QAM transmissions but with different resulting SNR. The 16 QAM results are shown in Fig. 9 for different numbers of receive elements. When the maximum transmit voltage was 500 mV and above, there were no errors in received data, which implies that the BER was lower than 7.4e-6 with data rate of 3 Mbps.

The BER performance of 256 QAM is shown in Fig. 10. A BER of 1.8e-5 was achieved when the transmit voltage was 1 V peak with a data rate of 6 Mbps. In terms of BER, the 256 QAM performance was not as good as 16 QAM because we doubled the bits for 256 QAM but the transmit power was unchanged. The Eb/N0 was half of that of 16 QAM. While the BER was not as good as the 16 QAM signals, the BER was still low and acceptable for video streaming with moderate forward error correction.

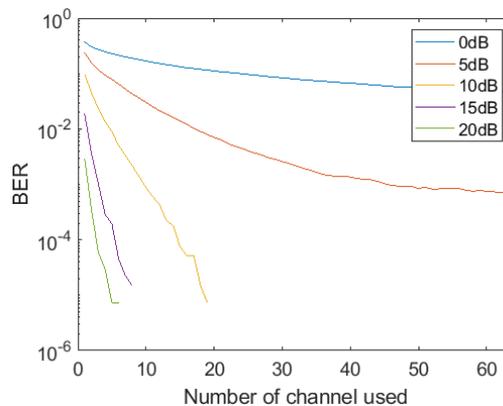

Fig. 8. BER results of 16 QAM simulation with measured phantom FIR.

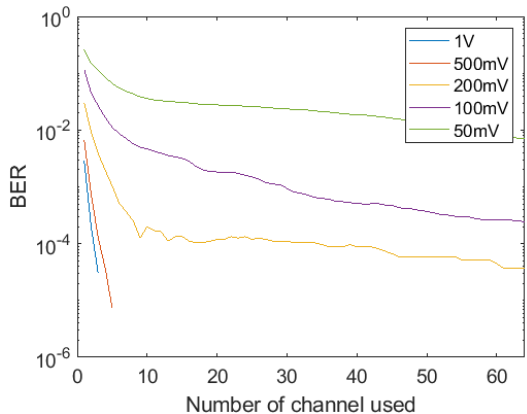

Fig.9. BER performance for 16 QAM through the phantom.

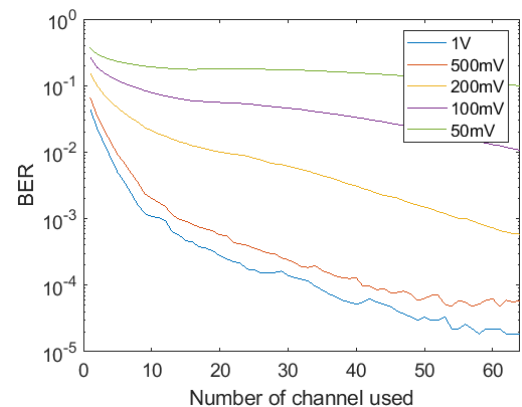

Fig.10. BER performance for 256 QAM through the phantom.

The BER performance for the high bandwidth mode, which used a 2.3 MHz bandwidth, is shown in Fig. 11. A BER of 2e-

4 was achieved when the transmit voltage was 1 V peak with a corresponding data rate of 15 Mbps.

The constellation diagrams of both 16 QAM and 256 QAM at 1 V peak transmit voltage are also presented. Figure 12 shows the constellation diagrams when using a single channel for receive and 64 channels MRC. In 16 QAM mode a data rate of 3 Mbps was achieved using 64 channels. The average error vector magnitude (EVM) decreased from -20.8 dB to -33.1 dB as more channels were used.

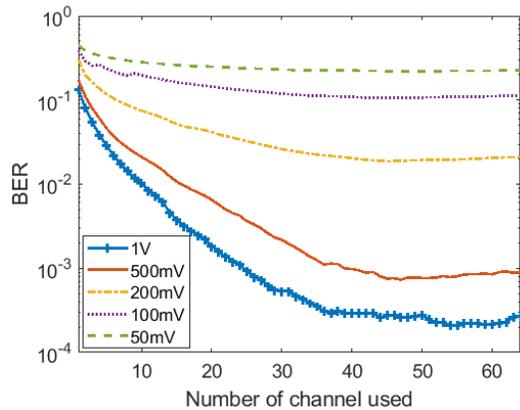

Fig.11. BER performance for high bandwidth 256 QAM through the phantom.

Figure 13 shows the constellation diagrams for a single receive channel and 64 receive channels when using 256 QAM. Using 256 QAM mode a data rate of 6 Mbps was achieved. The average EVM decreased from -23 dB to -33.3dB as the channel count increased.

Figure 14 shows the constellation diagrams when using a single receive channel and 64 receive channels and MRC in the high bandwidth mode. Using the high bandwidth mode with 256 QAM, a data rate of 15 Mbps was achieved. The average EVM decreased from -17.2 dB to -32.2 dB as the channel count increased to 64 elements.

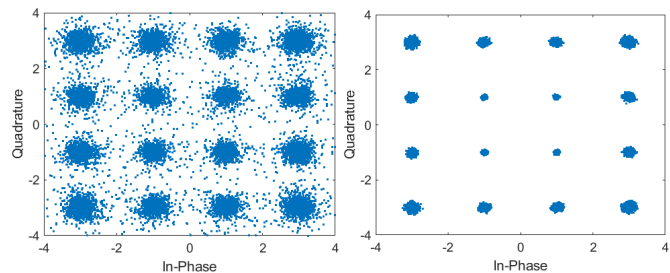

Fig.12. 16 QAM constellation diagram when using a single receive channel (left) and using 64 receive channels and MRC (right).

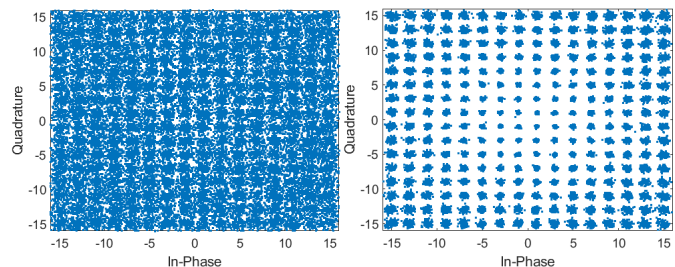

Fig.13. 256 QAM constellation diagram when using a single receive channel (left) and using 64 receive channels and MRC (right).

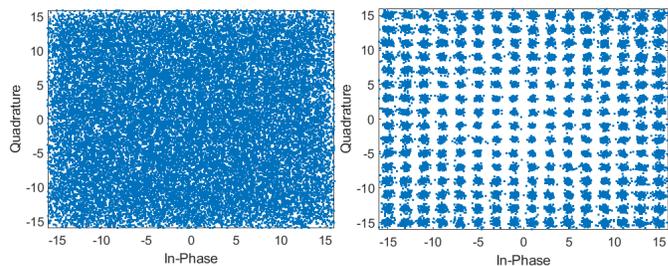

Fig.14. High bandwidth 256 QAM constellation diagram when using a single receive channel (left) and using 64 receive channels and MRC (right).

We tested the real video transmission with the FPGA design through the phantom. Figure 15 shows an image frame received with IP103 probe through the phantom. The image had 160*50 resolution in RGB565 format (16 bits per pixel). The maximum frame rate was 0.5 frame per second when using 10 channels and MRC mode. Performance was limited by the processing speed of host computer. If not limited by the receiver, the FPGA could transmit 24 frames per second, at a 3 Mbps data rate.

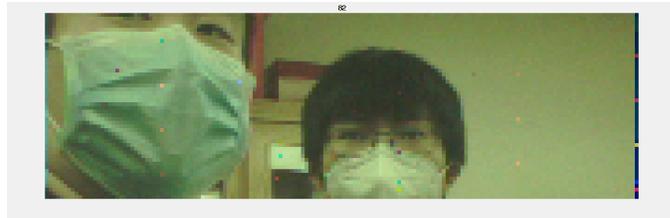

Fig.15. Sample image from video transmitted through the phantom using 16 QAM.

A summary of performance for each operation mode is listed in Table III. Using the array system, we were able to use a small microcrystal with lower power output and still achieve data rates of 15 Mbps through 6.5 cm of tissue-mimicking material.

### B. In vivo Results

Figure 16 shows an image frame received when streaming video signals with ultrasound through the abdomen of a rabbit *in vivo* using a C5-2 curvilinear array probe connected to the Verasonics system. The transmitting crystal was embedded at approximately 2 cm inside the rabbit abdomen. Due to the processing speed limitation of the Verasonics system, we set the number of channels used for MRC to 10 for the image transmission at data rate of 3 Mbps. The frame rate and resolution were the same as in the phantom.

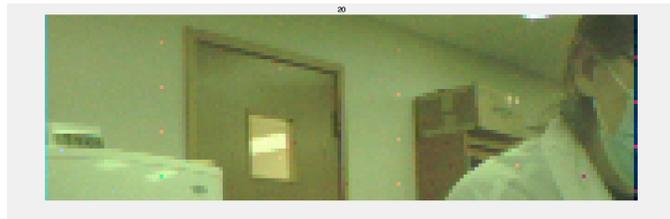

Fig. 16. Sample image from video transmitted through rabbit abdomen using 16 QAM.

## IV. DISCUSSION

In this study, we designed an OFDM modulation scheme, which can use ultrasound as a communication channel to transfer signals at HD video rates. The transmission signals were constructed on an FPGA board to enable real-time video transmission on a small platform. The performance of the communication scheme indicated that the communication protocol was suitable for video capable ultrasound capsule endoscopy. The FPGA transmitted signals at low power while still achieving data rates of up to 15 Mbps.

TABLE III
PERFORMANCE SUMMARY (BER)

| Modulation | 16 QAM | | 256 QAM | | 256 QAM | |
|---|---|---|---|---|---|---|
| Bandwidth | 937.5 kHz | | 937.5 kHz | | 2.343 MHz | |
| Data Rate | 3 Mbps | | 6 Mbps | | 15 Mbps | |
| Voltage | Single | MRC | Single | MRC | Single | MRC |
| 1V | 2.619e-3 | <7.398e-6 | 0.099 | 1.850e-5 | 0.133 | 2.027e-4 |
| 500 mV | 5.526e-3 | <7.398e-6 | 0.066 | 5.919e-5 | 0.172 | 7.287e-4 |
| 200 mV | 2.421e-2 | <7.398e-6 | 0.176 | 5.845e-4 | 0.302 | 1.870e-2 |
| 100 mV | 8.000e-2 | 2.959e-5 | 0.284 | 0.011 | 0.391 | 0.101 |
| 50 mV | 0.227 | 3.033e-3 | 0.376 | 0.010 | 0.445 | 0.220 |

TABLE IV
PERFORMANCE COMPARISON

| | [1] | [15] | [11] | This work |
|---|---|---|---|---|
| Medium | Pork loin | Phantom | beef | Phantom |
| Transducer | Focus-Focus | Focus-Focus | Single-Single | Single-Array |
| Distance | 5.86 cm | 10 cm | 10 cm | 6.5 cm |
| Modulation | 64 QAM | OFDM+8 QAM | OFDM+16 QAM | OFDM+256 QAM |
| Data Rate | 30 Mbps | 12 Mbps | 340 kbps | 6 Mbps/15 Mbps |
| BER | <1e-4 | 1.9e-4 | <1e-4 | 1.8e-5/2e-4 |

The use of an array at the receiving end allowed for improved transmission rates over single transducer receivers. Using a single channel on the array resulted in low SNR, which translated to lower data rates to maintain low BER. However, by combining the data from multiple channels of the array, the SNR was increased allowing data rates of up to 15 Mbps with BER of 2e-4. Therefore, the study demonstrates the importance of the receiving hardware in improving signal detection and conversion when using small ultrasonic sources, such as might be used in small implantable medical devices. Furthermore, the array reduced the need to perfectly align the source and receiver. In this case a 1D array was used allowing focusing and alignment along the lateral dimension.

These high data rates were achieved through a tissue-mimicking phantom, which was constructed to model what might be observed from transmission through the human abdomen. *In vivo* experiments in a rabbit abdomen also demonstrated the ability to transmit video using ultrasound as a communication medium with data rates of 3 Mbps. Table IV provides a comparison of the data rate, BER and system parameters between the setup provided here and the results from other studies. Our results demonstrate high data rates (>3 Mbps) even using unfocused microcrystals as transmitters.

The real-time video transmission based on the current platform is still limited by the processing ability of the software-based receiver. As the active receive channel number increased, the frame rate dropped rapidly. As the number of active receive channels increased, we needed to perform DDC and FFT on every active receiving channel, which means the number of filters and complex multipliers were linearly increasing with the number of active receive channels. This multiple channel calculation can be parallelized for speed up allowing real-time video. For the next step, we will design a hardware based multichannel receiver that moves all the receive processing, especially DDC and FFT, to an FPGA. We will also implement channel coding and video coding to further improve performance.

To create a capsule endoscopy system requires design of both the receive hardware and the transmit hardware. The elevational focus of the curvilinear or linear arrays used as a receiver in the work limited the positioning of the receiver with the transmitter, i.e., the transmitter had to be within the elevational plane of the array. Our future work includes a 2D array sitting on the surface of the abdomen constructed specifically for receiving signals from a small source with low directionality moving throughout the small intestines (i.e., a capsule endoscopy pill with a small microcrystal ultrasonic source). The 2D array would allow source localization of the transmitter over a volume and an MRC procedure to focus the receive pattern and increase SNR.

Furthermore, to move toward capsule endoscopy, the form factor of the transmit hardware needs to be minimized to fit the size of capsule. While camera and ultrasonic transducer options exist that would fit in an endoscopy capsule, to our knowledge at this time off-the-shelf processing hardware solutions are not available that fit the required footprint. One solution is to design a mixed signal System-on-Chip (SoC) that incorporates video encoding, channel encoding, OFDM modulation and DAC on a single chip. SoC designs and hardware have been constructed for this purpose [19]. The power supply, which is limited by the onboard battery, is also a challenge. A highly integrated SoC can lower the power consumption compared to a separate FPGA and DAC architecture as we currently use allowing for longer operation of an ultrasound-based capsule endoscopy.

## ACKNOWLEDGEMENTS

This work was supported by a grant from the National Institutes of Health (NIH R21EB025327).